# Polarization rotation and the electrocaloric effect in barium titanate

H. H. Wu[1] and R. E. Cohen[1,2*]

[1]Department of Earth and Environmental Sciences, Ludwig-Maximilians-Universität, Munich 80333, Germany

[2]Extreme Materials Initiative, Geophysical Laboratory, Carnegie Institution for Science, Washington, DC 20015-1305, United States

**Abstract**

We study the electrocaloric effect in the classic ferroelectric $BaTiO_3$ through a series of phase transitions driven by applied electric field and temperature. We find both negative and positive electrocaloric effects, with the negative electrocaloric effect, where temperature decreases with applied field, in monoclinic phases. Macroscopic polarization rotation is evident through the monoclinic and orthorhombic phases under applied field, and is responsible for the negative electrocaloric effect.

[*]Corresponding author. E-mail address: rcohen@carnegiescience.edu

## 1. Introduction

The electrocaloric (EC) effect is the temperature change of a material in response to an applied electric field, and has been proposed as an efficient form of the solid state refrigeration [1-4]. The strongest EC effect typically is observed in ferroelectric materials near a ferroelectric phase transition, where application of a moderate electric field would result in a large change of the electric polarization. The EC effect in $BaTiO_3$ has been studied both experimentally [5-9] and theoretically [10-13]. Most previous work focused on the temperature region above room temperature, and considered only the ferroelectric-paraelectric phase transition. However, large polarization changes also occur between the ferroelectric-ferroelectric phase transitions. Polarization rotation induced by an external electric field is related to phase transition behavior and many important properties of ferroelectric materials [14-21]. For example, with a first principles study of the ferroelectric perovskite $BaTiO_3$ [17], Fu and Cohen found that a large piezoelectric response can be driven by polarization rotation induced by an external electric field. In PMN-PT there is a critical line under applied electric field, where rotating the polarization becomes extremely easy [18,22,23]. The negative EC effect has been observed in ferroelectric/antiferroelectric materials recently [24-33]. The polarization component increases at the phase transition temperature R→O and O→T under zero applied electric field, indicating the intrinsic existence of a negative EC effect in $BaTiO_3$ [34,35], thus inspiring us to study this classic system further.

A first-principles based effective Hamiltonian [36] gave promising results for the phase diagram of $BaTiO_3$ under applied field E, showing the phase boundaries for the rhombohedral ( R ), orthorhombic ( O ), and tetragonal ( T ) phases as functions of T and E. However, under applied field the polarization could rotate (Figure 1), giving also $M_A$ and $M_B$, as have been predicted and observed in $Pb(Zr_{1-x}Ti_x)O3$ [37], $Pb(Zn_{1/3}Nb_{2/3})$-$PbTiO_3$ [38,39] and in $Pb(Mg_{1/3}Nb_{2/3})O_3$-$PbTiO_3$ [23,40-45], and have been considered within an extended Devonshire model [46]. Despite the electric field induced monoclinic phases in barium titanate have also been observed by experiments [47] and numerical calculations [48], detailed studies and the relationship between the monoclinic phases and the EC effect is still lacking. A first-principles based shell model, fit to density functional theory total energies using the PBEsol exchange correlation functional [49], is promising and requires no assumptions about the important decreases of freedom, and includes the effects of hard-modes as well as the soft-modes in computing thermal behavior, unlike effective Hamiltonians.

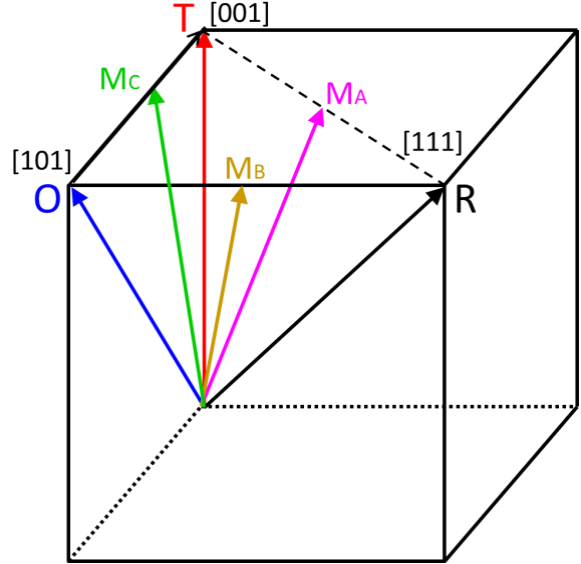

Figure 1. Definition of different phases and possible phases transition paths with respect to temperature, electric field and pressure , The monoclinic $M_A$ $(P_1 = P_2 < P_3)$, $M_B$ $(P_2 < P_1 = P_3)$, $M_C$ phase $(P_3 > P_1 \neq 0, P_2 = 0)$, rhombohedral (R) $(P_3 = P_1 = P_2 > 0)$, orthorhombic (O) $(P_1 = P_3, P_2 = 0)$, and tetragonal (T) $(P_3 > 0, P_1 = P_2 = 0)$ phases are indicated. The arrows represent the polarization directions. The rotation paths would be between these directions.

## 2. Model and computational details

We performed classical molecular dynamics (MD) with DL-POLY [50] using this first principles based shell model within the constant (N, σ, T) ensemble. We use $12 \times 12 \times 12$ supercells (8640 atoms) under periodic boundary conditions. The thermostat and barostat relaxation times are set to 0.25 ps and 0.35 ps, respectively. All shells are assigned a mass of 2 a.u. [51]. Our MD runs consist of at least 100 000 time steps, with data collection after 60 000 time steps, with a time step of 0.4 fs, giving run times of at least 40 ps. We found it important to pole the simulated systems before measurement runs; we poled them along [001] with electric field strength of 200 MV/m. We determined the polarization at each time step by comparing the atomic positions to those of perfect cubic perovskite [52] . It is well known that in $BaTiO_3$, the phase transitions are largely order-disorder in character, and even the cubic paraelectric phase has atomic displacements, which are obvious from the streaking in X-ray diffraction, for example [53,54]. Here we do not consider the transition character, but only the bulk average polarization vector as a function of applied field and temperature.

We find the EC effect using the Maxwell relation [55], that gives the adiabatic temperature change (ATC) $\Delta T$, as the applied electric field increases from an initial value to a final value $E$, is calculated from

$$\Delta T = -\int_{E_a}^{E_b} \frac{TV}{C_{p,E}} \left(\frac{\partial \boldsymbol{P}}{\partial T}\right)_E \cdot d\boldsymbol{E}$$

where $C_{p,E}$ is the heat capacity of the supercell [22], $P$ is the macroscopic polarization, $E$ is the applied electric field, $V$ is the volume of the supercell, $T$ is the absolute temperature, and $p$ is the pressure.

## 3. Results

*3.1. Macroscopic properties and phase diagram*

Without external electric field and at zero pressure, we find the expected phase transition sequence with temperature increase of R→O→T→C, (Fig. 2[a1]-[b1]), as is well known from experiments [1,34]. Under applied electric field $E_3$=50 MV/m (Figure 2[b1]-[b2]), the polarization components $P_1$ and $P_2$ are equal at low temperatures. The lattice parameters also are consistent with a monoclinic $M_A$ structure. With temperature increase, another monoclinic $M_C$ phase occurs. With temperature further increase to 250K, the phase transition from $M_C$ to $T$ phase appears. When the external electric field increases to $E_3$ =200 MV/m (Figure 2[c1]-[c2]), the phase transition temperatures of $M_A$ to $M_C$ and $M_C$ to T decreases. The increase of polarization component $P_3$ along the external electric field direction is due to polarization rotation.

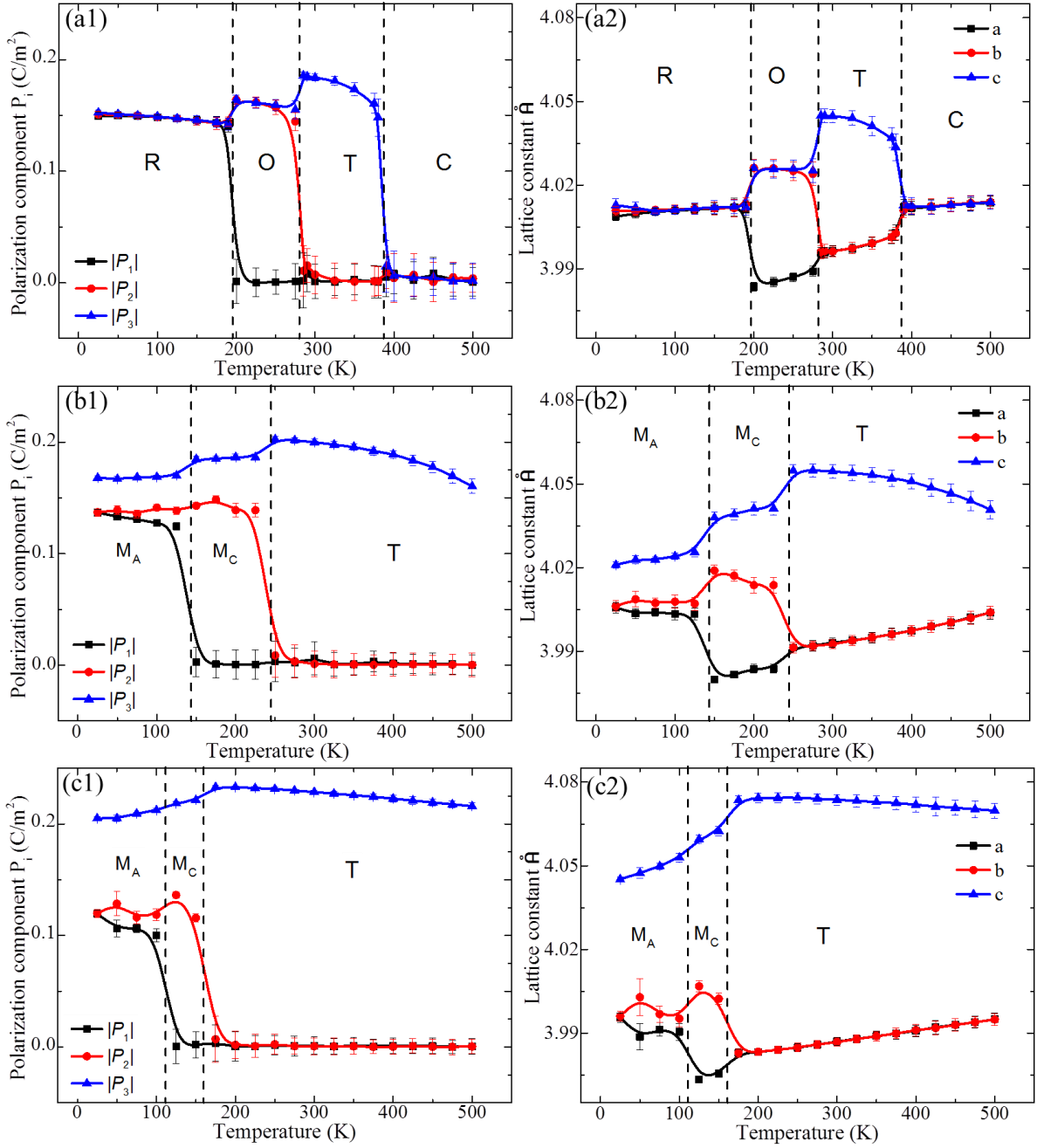

Figure 2. Average polarization components (a1)-(c1), and lattice parameters (a2)-(c2) as functions of temperature under zero pressure. (a1) and (a2) correspond to external electric field $E_3$=0 MV/m, (b1) and (b2) correspond to external electric field $E_3$=50 MV/m, and (c1) and (c2) corresponds to external electric field $E_3$=200 MV/m.

Under a hydrostatic pressure 4GPa (Figure 3), the phase transition paths with temperature increase are the same as that under zero pressure. However, the phase transition temperatures are reduced to lower temperatures.

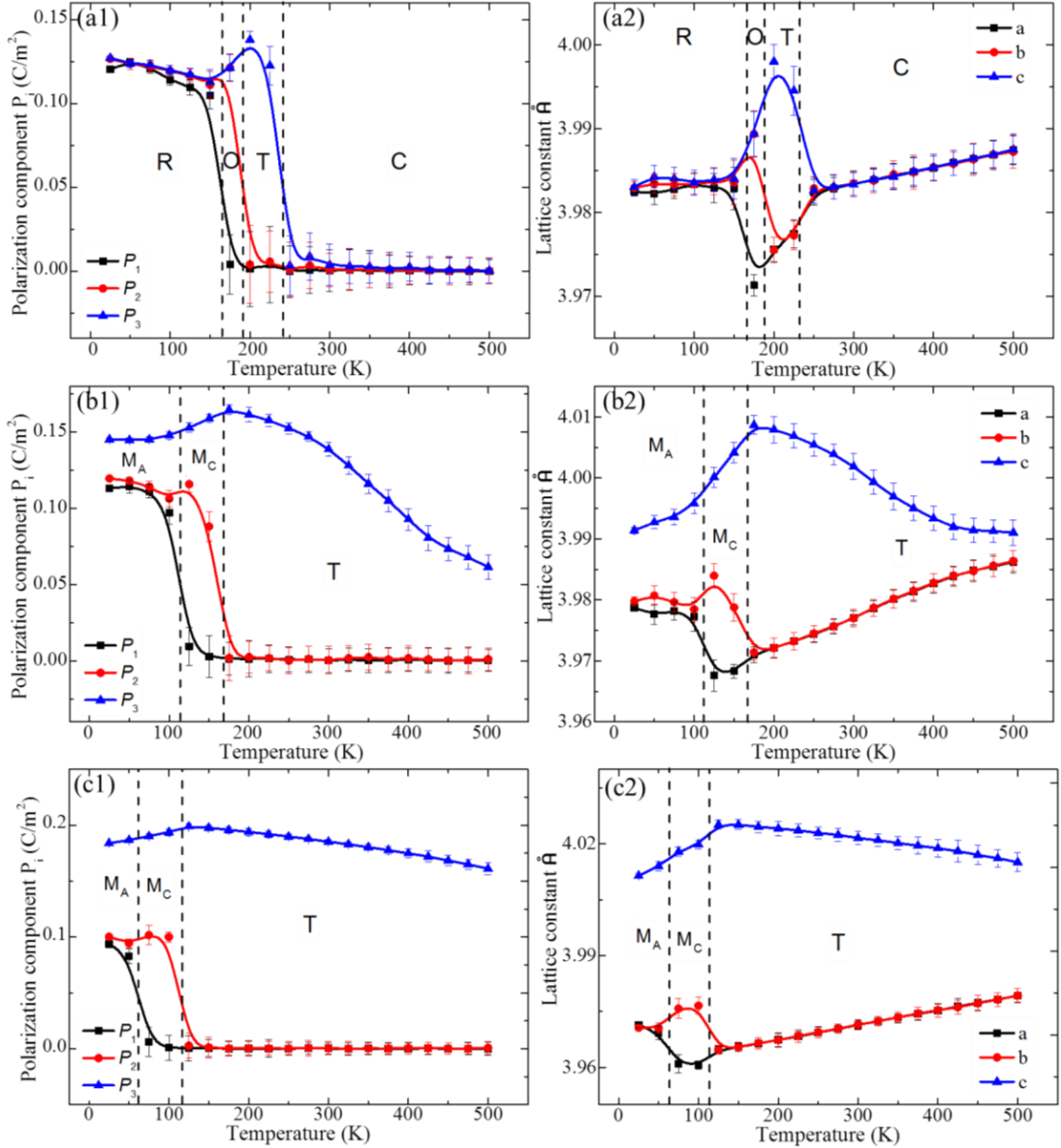

Figure 3. Average polarization components (a1)-(c1), and lattice parameters (a2)-(c2) as functions of temperature under pressure of 4GPa. (a1) and (a2) correspond to external electric field $E_3$=0 MV/m, (b1) and (b2) correspond to external electric field $E_3$=50 MV/m, and (c1) and (c2) correspond to the external electric field $E_3$=200 MV/m.

From these results, we build the electric field–temperature phase diagram (Figure 4). Solid line corresponds to zero pressure case, and dash line corresponds to the case under a hydrostatic pressure 4GPa. The phase transition temperature of ferroelectric materials is a function of external electric field. For example, with electric field increase, the phase transition temperature of $M_A \rightarrow M_C$ and $M_C \rightarrow T$ decreases monotonically. The slope of the applied electric field versus the phase transition temperature $\Delta E/\Delta T$ under zero pressure are basically -2.5 KV/(m·K) and -1.6 KV/(m·K), respectively. Such a dependence of phase transition temperature on external electric field has also been observed in experiments [18,55] and numerical calculations [23,48].

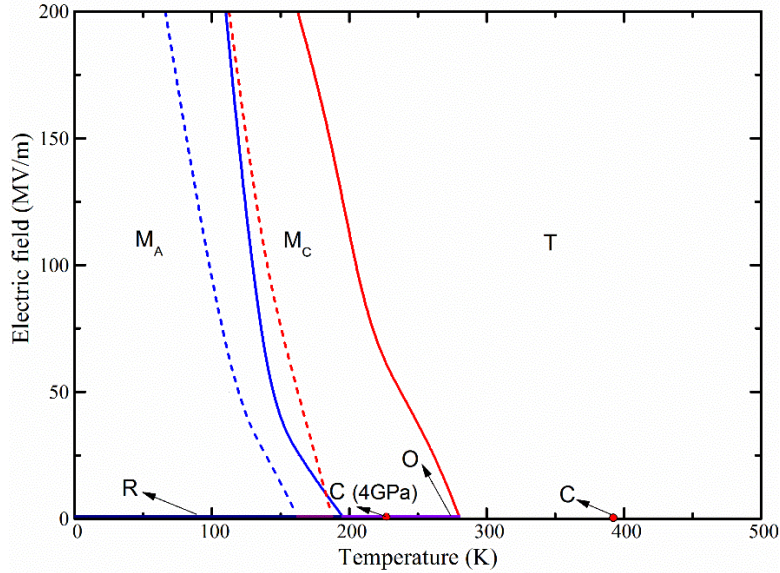

Figure 4. Phase diagram with respect to different electric fields and temperature for barium titanate. The solid line corresponds to zero pressure case, and dash line corresponds to the case under a hydrostatic pressure 4GPa.

*3.2. Polarization rotation mechanism*

The trend of polarization component $P_3$ versus temperature changes with applied electric field (Figure 5[a]). Take zero electric field as an example, the polarization component $P_3$ decreases with temperature increase when temperature is lower than 175 K, whereas the polarization $P_3$ jumps to a higher value at the phase transition temperature from R to O and also from O to T, whereas the zero-field total polarization $P_t$ (Figure 5[b]) always decrease with temperature increase in the entire temperature range from 25K to 525K. The maximum value of polarization component $P_3$ under electric field varying from 0 to 200 MV/m shifts from 300K to 175K, corresponding to the phase boundary from $M_C$ phase to T phase (Figure 4).

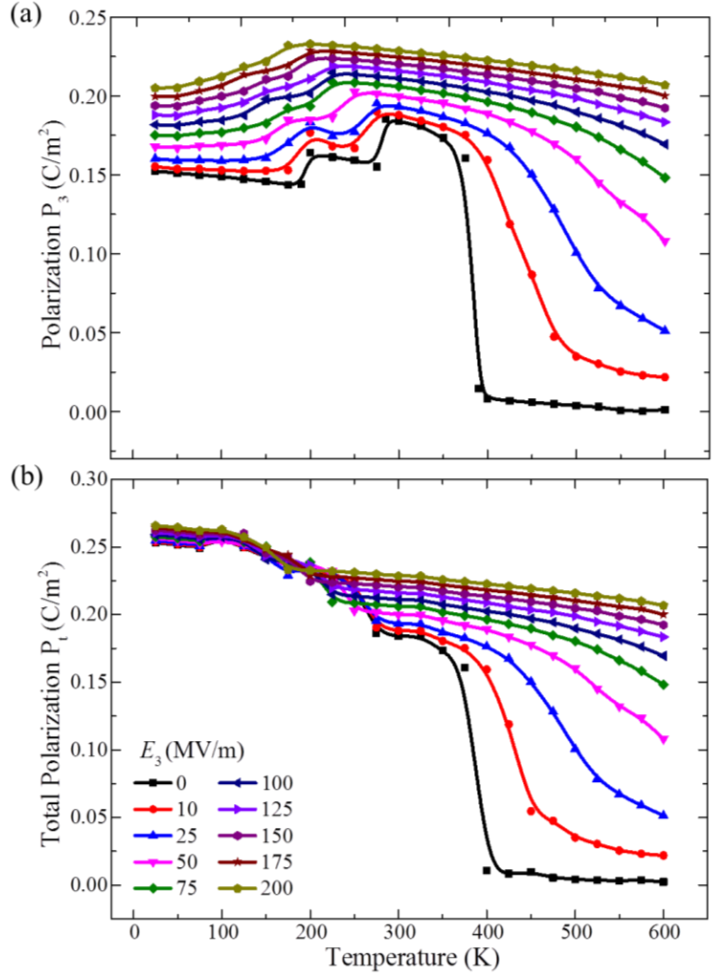

Figure 5. Simulation results under zero pressure. (a) Average polarization component $P_3$ and (b) average total polarization $P_t$ as functions of temperature and electric field.

Under a hydrostatic pressure 4 GPa, the trend of polarization component $P_3$ versus temperature also changes non-monotonically with the applied electric field $E_3$ increase (Figure 6[a]). For example, without electric field, the polarization component $P_3$ decreases in the temperature range 25K to 150K, and then it increases when the temperature increases to 200K. However, with the temperature further increase from 200K to 300K, the polarization component $P_3$ decrease to zero. The total polarization in Figure 6(b) decreases with temperature increase for all of the external electric fields. Combined with the phase diagram of Figure 4, the increase of polarization $P_3$ with temperature increase occurs in monoclinic phase and reaches a maximum polarization $P_3$ exactly at the phase transition temperature from $M_C$ to T, which further prove that the increase of polarization component $P_3$ with temperature increase is caused by the polarization rotation.

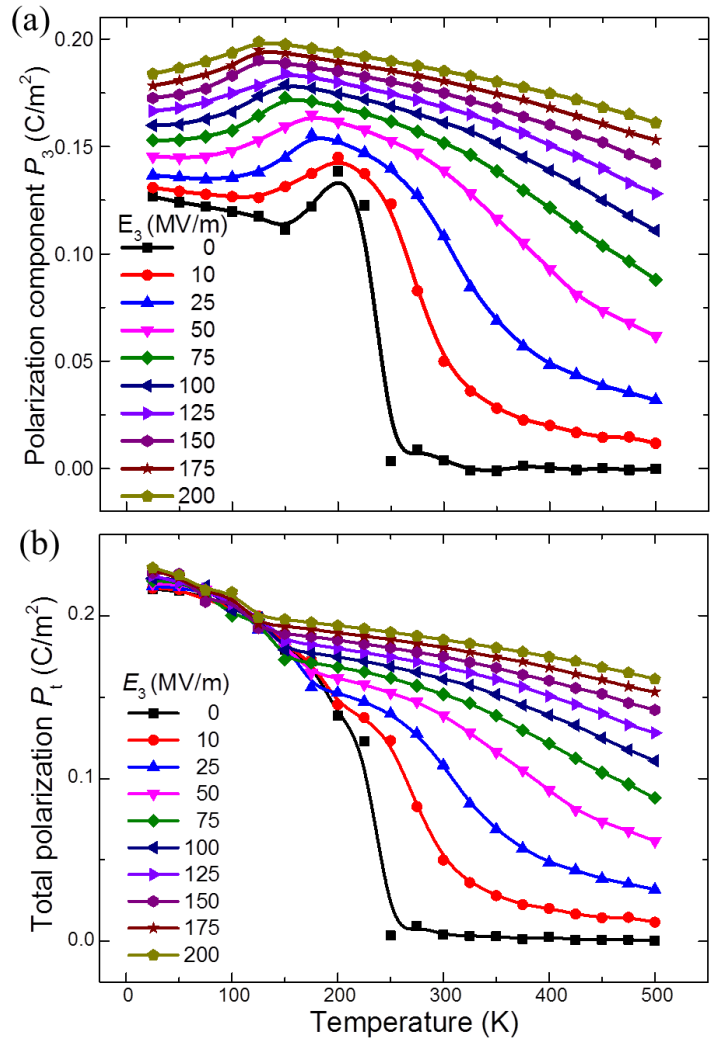

Figure 6. Simulation results under 4 GPa pressure. (a) Average polarization component $P_3$ and (b) average total polarization $P_t$ as functions of temperature and electric field.

*3.3. Electrocaloric effect*

The increase of polarization component $P_3$ along the applied electric field $E_3$ indicates the presence of negative EC effect, and we observe negative EC effect in the temperature range from 50K to 300K (Figure 7[a]). Interestingly, there are two peaks of the negative EC effect under low electric field, which correspond to the two phase transitions $M_A \rightarrow M_C$ and $M_C \rightarrow T$. The polarization component $P_3$ always decreases with temperature increase above 300K, so that the EC effect above 300K is positive. Moreover, due to the significant polarization change near the phase transition temperature, where the EC effect reaches its peak value under electric field, the peak of the positive EC effect shifts to higher temperature with increase in external electric field. At zero pressure the peak of positive ATC ΔT at 10 MV/m is at 400K, whereas at 200MV/m, the peak ATC ΔT occurs at 525K. Under hydrostatic pressure 4GPa, only one negative ΔT peak is observed in low temperatures (Figure 7[b]). The temperature of negative peak position shifts significantly toward lower temperatures, accompanied by an increase in magnitude. Above the temperature 200K, the polarization component $P_3$ decrease with temperature increase in all simulated electric fields, therefore, the EC effect above 200K are positive. Under hydrostatic pressure 4GPa, the polarization component $P_3$ becomes much smoother than that under zero hydrostatic pressure (Figure 6[a]), which causes the ATC ΔT to be more diffuse and the temperature stability at high temperature. Since the negative EC effect occurs within the electric field induced monoclinic phases, to measure the negative EC effect of $BaTiO_3$ in experiments, the temperature region should be lower than the phase transition temperature O→T. Moreover, the applied [001] electric field should be large enough to induce the monoclinic phase from R or O phase.

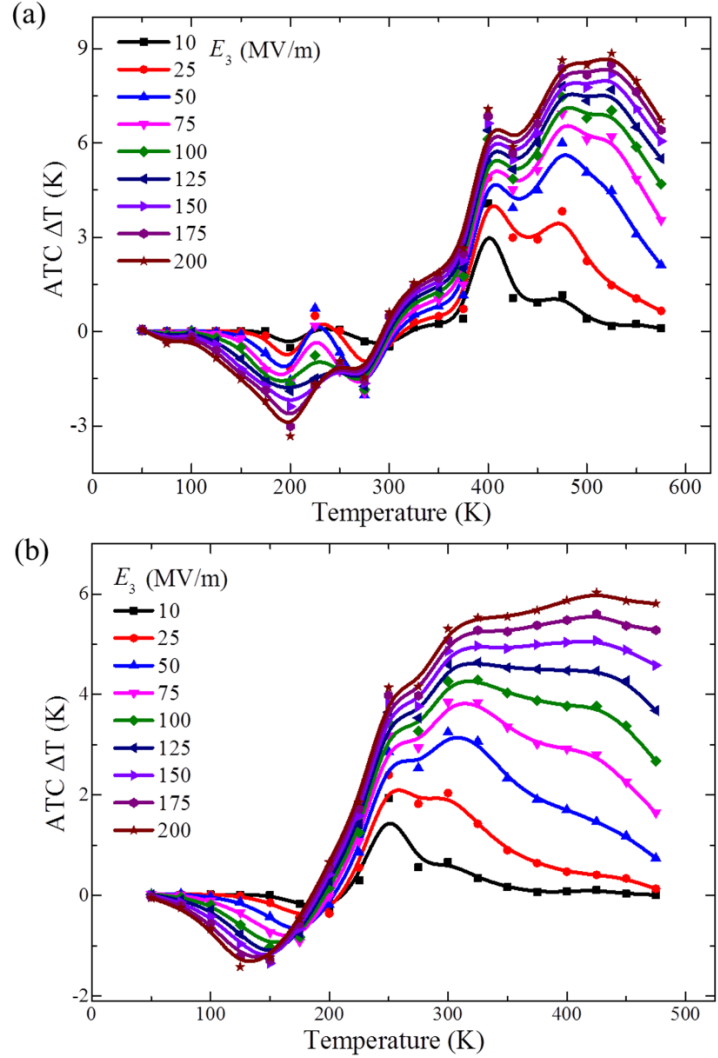

Figure 7. Adiabatic temperature change (ATC) $\Delta T$ of BaTiO$_3$ as a function of temperature and electric field at (a) 0GPa, (b) 4GPa.

The Joule heating from leakage current is harmful for a highly efficient EC cooling of ferroelectric materials. Therefore, the samples should be carefully fabricated in experiments. Generally, the Joule heating from leakage current can be ignored for high quality samples [56,57], i.e., the EC effect in antiferroelectric PbZrO$_3$ thin films studied by Parui and Krupanidhi [56]; the leakage current measurement at 235 ℃ under 51 MV/m shows ~6 μA as an upper bound of steady-state leakage current. The Joule heating of this magnitude can increase the temperature by ~0.1 K over one quarter of cycle, which is negligible compared to the peak EC effect of ~5.2 K.

## 4. Discussion and conclusions

In the current work, the calculated zero-field phase transition temperatures R→O (195K), O→T (280K) and T→C (385K) are in better agreement with experimental values [58,59] (203K, 278K, and 393K) than previous results [49] (180K, 250 K, and 340 K) reported by Vielma and Schneider. The major reason is the use of poling in our simulations, as described above, and the relatively long simulation times here. We performed classical MD, and so we use the classical Dulong-Petit value of molar heat capacity in calculation of the EC effects, so the EC effect is underestimated below the Debye temperature, 430K [61]. For example, the heat capacities of $BaTiO_3$ over a wide temperature range were studied by using experimental ac-hot probe method [60], an extended rigid ion model [61] and quasi-harmonic Debye model [62]; all those observations found that the heat capacity of barium titanate at sufficient high temperature, i.e. above the Debye temperature, does not depend much on temperature and tends to approach the classical limit 124.7 J mol$^{-1}$ K$^{-1}$. However, the heat capacity decrease non-linearly below the Debye temperature, i.e., the heat capacity[61] at 100K is 53 J mol$^{-1}$ K$^{-1}$, less than half of that at Debye temperature. In this case, the real EC effect magnitude will be ΔT=-0.58 K under $\Delta E_3$=200 MV/m, which is 2.35 times larger than the EC effect by employing the classical molar heat capacity. We find that the polarization rotation plays an important role in the monoclinic phases in our work, and thus leads to the polarization increase and negative EC effect. All the phase structures occur in our simulation does not make any prior assumptions about transient microstructures, where phase transition is a direct consequence of the minimization process of the total free energy over the entire simulated system. Previously, different mechanisms for the negative EC effect have been proposed, such as the noncollinearity between the electric field and the polarization [27], a pseudo-first-order phase transition [24]. For example, Ponomareva and Lisenkov attributed the occurrence of the negative EC effect to the noncollinearity between polarization and electric field. However, noncollinearity of applied electric field and polarization is not always the sufficient condition for occurrence of negative EC effect. The polarization increase with temperature (negative EC effect) can only occur when the external electric field is strong enough to induce the polarization rotation. For instance, when the external electric field $E_3$=10 MV/m is applied along [001] direction within the monoclinic $M_C$ phase (Figure 3), the polarization component $P_3$ of Figure 4(a) decrease with temperature increase, and thus $\Delta E_3$=10 MV/m of Figure 5(a) is positive in the temperature range 200-250K. Therefore, the proposed mechanism of polarization rotation is more universal for explaining the occurrence of the negative EC effect in all ferroelectrics, relaxors, and antiferroelectrics.

## Acknowledgments

The authors gratefully acknowledge the Gauss Centre for Supercomputing e.V. (www.gauss-centre.eu) for funding this project by providing computing time on the GCS Supercomputer SuperMUC at Leibniz Supercomputing Centre (LRZ, www.lrz.de). This work was supported by the European Research Council under the Advanced Grant ToMCaT (Theory of Mantle, Core, and Technological Materials) and by the Carnegie Institution for Science. The authors appreciate their fruitful discussions with Shi Liu and Yangzheng Lin.